\documentstyle[12pt]{l-aa}

\def\e20{\times 10^{20} {\rm cm}^{-2} }
\def\sax{{\it BeppoSAX}}
\def\NH{N_{\rm H}}
\def\ecs{\rm \,erg~ cm^{-2}\,s^{-1} }
\def\mincir{\ \raise -2.truept\hbox{\rlap{\hbox{$\sim$}}\raise5.truept  
\hbox{$<$}\ }}                        

\begin{document}

\thesaurus{03(11.02.1; 13.25.3)}	

\input psfig.tex

\title{ \sax\ Spectral Survey of soft X-ray selected BL Lacs}

\author{A. Wolter,
\inst{1} 
A. Comastri, 
\inst{2}
G. Ghisellini,
\inst{1}
P. Giommi,
\inst{3}
M. Guainazzi,
\inst{3}
 T. Maccacaro, 
\inst{1}
L. Maraschi,
\inst{1}
P. Padovani, 
\inst{4,5,6}
C.M. Raiteri, 
\inst{7}
G. Tagliaferri,
\inst{1}
C.M. Urry, 
\inst{4} 
M. Villata.
\inst{7}
}

\offprints{A. Wolter: anna@brera.mi.astro.it}

\institute{Osservatorio Astronomico di Brera
		Via Brera, 28
		20121 MILANO, Italy
\and Osservatorio Astronomico di Bologna, Via Zamboni 33, 40126 Bologna, Italy.
\and A.S.I., Beppo-SAX Science Data Center, Via Corcolle 19, I-00131, Roma, Italy
\and Space Telescope Science Institute, 3700 San Martin Drive, Baltimore, MD. 
21218, USA
\and Affiliated to the Astrophysics Division, Space Science Department, 
European Space Agency
\and On leave from Dipartimento di Fisica, II Universit\`a di Roma ``Tor Vergata", Italy
\and Osservatorio Astronomico di Torino, Strada Osservatorio 20, I-10025 Pino Torinese (TO), Italy}

\date{Received ....; accepted ....}
\maketitle
\markboth{A. Wolter et al.}{\sax\ spectral survey of soft X-ray selected BL Lacs}

\begin{abstract}
We present X-ray spectra obtained with \sax\ (Satellite per Astronomia X)
of 10 BL Lac objects,
selected from the Einstein Medium Sensitivity and Slew Surveys. 
We find that in about half of the  objects a fit in the 
0.1-10 keV range with a single power law and free absorption yields 
values of $N_{\rm H}$ larger than the Galactic ones.  
In most of these cases, however, broken power law fits with 
$N_{\rm H}$ fixed at the Galactic values yield an alternative,
better description of the data and indicate a steepening of the spectrum 
with increasing energy. 
One object (1ES1101-232) is detected up to $\sim$ 100 keV.
Its spectral energy distribution (SED) peaks in the medium energy X-ray band.
For each object we compute the peak frequency of the SED from 
multifrequency data.
The spectral indices $\alpha_x$ in the 2-10 keV band 
($F_\nu \propto \nu^{-\alpha_x}$) are smaller (i.e. flatter spectrum) 
for objects with higher peak frequencies. 
We therefore confirm and extend to higher energies the behavior 
already known for X-ray selected BL Lac objects in the ROSAT band.
We do not find spectral indices smaller than 1; however, the flat distribution 
of $\alpha_x$ and the correlation between $\alpha_x$ and peak frequency found 
from our data suggest that a number of objects may exist, which in the
quiescent status  have flatter spectrum and peak 
frequency in the hard X-ray range.

\end{abstract}

\section{Introduction}
\noindent
BL Lacertae objects are a rare type of Active Galactic Nuclei (AGN) 
characterized by strong and variable emission of non-thermal radiation 
across the entire electromagnetic spectrum, from radio waves to high 
energy $\gamma$-rays. 
In three cases (Mkn 421: Punch et al. 1992; Mkn 501: Quinn et al. 1996;
1ES 2344+514: Catanese et al., 1998)
the emission has been detected up to TeV energies.
BL Lacertae objects comprise  the most violent
(highly and rapidly variable, highly polarized) and most elusive 
(extremely difficult to find in optical surveys) sources amongst AGN.
Unlike most other AGN they do not show evidence (by definition) for 
strong emission lines  or large Infra-Red or UV excesses. 
The emission from radio to $\gamma$-rays
can be explained as due to synchrotron radiation up to a certain
maximum frequency (that ranges approximately from $10^{13}$ to  $10^{17}$ Hz), above
which a sharp turnover occurs until a second component due to Compton 
scattered radiation dominates, making these objects detectable up to the 
highest energies so far accessible (see e.g., Ulrich, Maraschi and Urry, 1997).
The extreme properties of BL Lacs require that the matter
emitting the radiation moves at relativistic speeds in the direction of the
observer. 

The spectral change from synchrotron to Compton radiation is crucial for
the understanding of the physics of BL Lacs. However, up to now this has 
been inferred only from the comparison of X-ray measurements carried out with 
different instruments and very often at different epochs. 
The wide energy band of \sax\ offers for the brightest objects the best 
opportunity 
to directly detect without ambiguity this spectral change and
to study the X-ray spectra at the same epoch over a large interval.

To this end we have undertaken a program
that aims at studying in detail the X-ray spectrum of a 
large and well defined subsample of soft X-ray selected BL Lacs. 
This sample
includes mostly objects that are expected to show strong 
spectral curvature and spectral breaks, since
the synchrotron break should occur just before or in the \sax\ band.
We aim at
measuring in detail the shape of the most energetic part of the synchrotron 
emission, and trying to establish where and how the Compton component 
becomes dominant. 
We also intend to look for the correlation between spectral slope 
and break energy found in ROSAT data (Padovani $\&$ Giommi, 1996;
Lamer, Brunner \& Staubert, 1996).

\section{The Sample of soft X-ray selected BL Lacs}

\begin{table*}
\caption{Journal of Observations}
\label{Jou}
\begin{tabular}[h]{| l c r c r c r|}
\hline
Name                  &z      &Obs. Date    &Exp. Time(sec) & net counts  & Exp.Time(sec) & net counts     \\
                      &       &d/m/y        &(LECS)         & (LECS)      & (MECS)        & (MECS)    	\\
\hline
MS0158.5+0019         &0.299  & 16-17/08/96 &4310 & 233.1 $\pm$17.5 &  12444 &  638.3 $\pm$26.5  \\
MS0317.0+1834         &0.19\phantom{0}  &    15/01/97 &4359 & 293.9 $\pm$19.0 &  14976 & 1905.3 $\pm$45.1  \\ 
1ES0347--121          &0.188  &    10/01/97 &6492 & 577.0 $\pm$26.9 &  10675 & 1254.9 $\pm$36.6  \\
1ES0414+009           &0.287  & 21-22/09/96 & --  & ------          &  11039 & 1858.4 $\pm$44.1  \\
1ES0502+675           & --    &   6-7/10/96 & --  & ------          &  11045 & 4117.3 $\pm$64.9  \\
MS0737.9+7441         &0.315  & 29-30/10/96 &3075 &  37.1 $\pm$7.8$^{\rm a}$ &  23279 &  735.9 $\pm$30.6  \\
1ES1101--232          &0.186  &     4/01/97 &5195 &2484.2 $\pm$51.1 &  13830 & 9509.3 $\pm$98.1  \\
1ES1133+704 (Mkn 180) &0.046  & 10-11/12/96 &4078 & 439.2 $\pm$22.6 &  18266 & 1940.7 $\pm$45.8  \\
MS1312.1--4221        &0.108  &    21/12/97 &3541 & 228.2 $\pm$17.5 &  \phantom{1}5555 &  446.7 $\pm$22.0  \\
1ES1517+656           & --    &     5/03/97 &4536 & 779.3 $\pm$29.5 &  11130 & 2273.3 $\pm$48.6  \\
            \hline
\end{tabular}
\begin{list}{}{}
\item[$^{\rm a}$] counts from a 3/4 of a circle, excluding the bottom-right
quadrant that contains a non-related source.
\end{list}
\end{table*}

Unlike any other type of AGNs,
more than 90\% of all known BL Lacs have been discovered either in radio or X-ray 
surveys. The former (often called RBLs) were found to 
show somewhat different properties from the latter
(often called XBLs). 
From
the viewpoint of the broad band spectrum the two classes differ mostly in
the position of the synchrotron break with RBL showing mostly a low energy
break and XBL showing the break at higher energies. 
A classification has recently been introduced where objects for which the 
break occurs at low energy (IR-Optical) are called LBLs (Low-frequency cut-off
BL Lacs) and objects where the turnover occurs at higher energies (UV-X-ray) 
are called HBLs (High-frequency cut-off BL Lacs; see Padovani \& Giommi, 1995). 
In this scheme most (but not all) RBL are
LBL and most (but not all) XBL are HBLs. 

At present, X-ray selected BL Lacs are mainly the result of surveys carried
out with the {\it Einstein} IPC. The Slew Survey sample (Perlman et al. 1996a)
covers essentially the entire high
Galactic latitude sky with a rather high flux limit, while the EMSS survey
(Gioia et al. 1990) is
deeper than the Slew Survey but covers only $\sim 800$ sq.deg. of the sky, 
with almost two orders of magnitude better sensitivity.
By selecting objects with flux [0.1--10 keV] higher than $10^{-11} \ecs $
in the Slew Survey and higher than $4 \times 10^{-12} \ecs$ in the EMSS
we obtain a sample that combines the advantages of a flux-limited 
(and therefore statistically well defined) sample with a wide coverage of 
the parameter space (i.e., X-ray and radio luminosity, redshift, $F_x/F_r$, 
etc.) which neither of the  two surveys alone would provide.

\noindent
The sample we present here includes the first 10 objects of this project
observed by the Narrow Field Instruments of \sax\ and represents
a significant fraction of the total sample that will be published
when available. Name(s) and redshift for each source are listed in
the Journal of Observations (Table~\ref{Jou}), described in the next Section.

\section{\sax\ Observations and Data Analysis}

The X-ray astronomy satellite \sax\ is a project of the Italian
Space Agency (ASI) with a participation of the Netherlands Agency for 
Aerospace Programs (NIVR).
The scientific payload comprises   four Narrow Field Instruments [NFI:
Low Energy Concentrator Spectrometer (LECS), Medium Energy Concentrator
Spectrometer (MECS), High Pressure Gas Scintillation Proportional Counter
(HPGSPC), and Phoswich Detector System (PDS)], all pointing in the
same direction, and two Wide Field Cameras (WFC), pointing in
opposite directions perpendicular to the NFI common axis. A detailed
description of the entire \sax\ mission can be found in Butler \& Scarsi (1990)
and Boella et al. (1997a). 

The MECS consists of three equal units, each composed of a grazing
incidence mirror unit and of a position sensitive gas scintillation 
proportional counter, with a field of view of
56 arcmin diameter, working range 1.3--10 keV, energy resolution $\sim 8\%$
and angular resolution $\sim 0.7$ arcmin (FWHM) at 6 keV.
The effective area at 6 keV is 155 cm$^2$ (Boella et al., 1997b)

The LECS is a unit similar to the MECS, with a thinner window that
grants a lower energy cut-off (sensitive in the energy range 0.1-10.0 keV)
but also reduces the FOV to 37 arcmin diameter (Parmar et al. 1997). 
The LECS energy resolution is a factor $\sim 2.4$ better
than that of the ROSAT PSPC, while the effective area is between 
a factor $\sim 6$ and 2 lower at 0.28 and 1.5 keV, respectively.

The PDS is a system of four crystals, sensitive in the 13--200~keV band and
mounted on a couple of rocking collimators, which points two units on the
targets and two units $3.5^{\circ}$ aside respectively, to monitor the
background.
The position of the collimators flips every 96 seconds.
Thanks to the stability of the instrumental background, the PDS has shown 
an unprecedented sensitivity in its energy range, allowing 3$\sigma$ detection 
of $\alpha \sim 1$ sources as faint as 10 mCrab with
10 ks of effective exposure time (Guainazzi \& Matteuzzi, 1997).

As of April 1997, 10 of the scheduled X-ray selected BL Lacs have been 
observed; the data have been preprocessed at the \sax\ SDC (Science
Data Center) and retrieved through the SDC archive. 
Table~\ref{Jou} lists a journal of the observations with the names
and the known redshifts for reference, the observation date and
the good-data times in the LECS and MECS, together with the extracted
net counts and photon counting statistics errors.
For two objects, 1ES0414+009 and 1ES0502+675, the LECS instrument
was not available during the observation. For the others, the LECS
exposure times are considerably reduced with respect to the MECS 
exposure times (a factor of 2 to 7) because the LECS can operate 
only when the spacecraft is not illuminated by the Sun.

\subsection{Variability}
Large intensity variations in the X-ray band are not uncommon for BL Lacs 
objects, and affect the entire spectral distribution; in a few, maybe
extreme, cases of the brightest HBL observed, even 
variations up to 2 orders of magnitude of the position of the emission peak in 
the spectral energy distribution have been detected
(e.g., Mkn 501: Pian et al., 1998; 
1ES 2344+514: Giommi, Padovani \& Perlman, 1998 and in preparation)

For this reason we have scrutinized all sources for variability over 
time scales of 500 or 1000 seconds, so as to have about ten different
bins in the course of the observation. No variability is observed 
during each observation (a fit with a constant is always acceptable).
Therefore we will analyze each dataset
as a whole to derive spectral information (see Section 3.2).

On time scales longer than the \sax\ pointing, variability is inferred
from comparison with other observations (see e.g., below in Section 6, and 
Figure~\ref{FigSeda}).
Most of the sources have been observed with ROSAT in 
an energy band very close to the \sax\ low energy band. We will therefore
present the ROSAT data and a comparison with the \sax\ ones
in Section 4.

\subsection{Spectral Analysis}

\begin{figure*}
\psfig{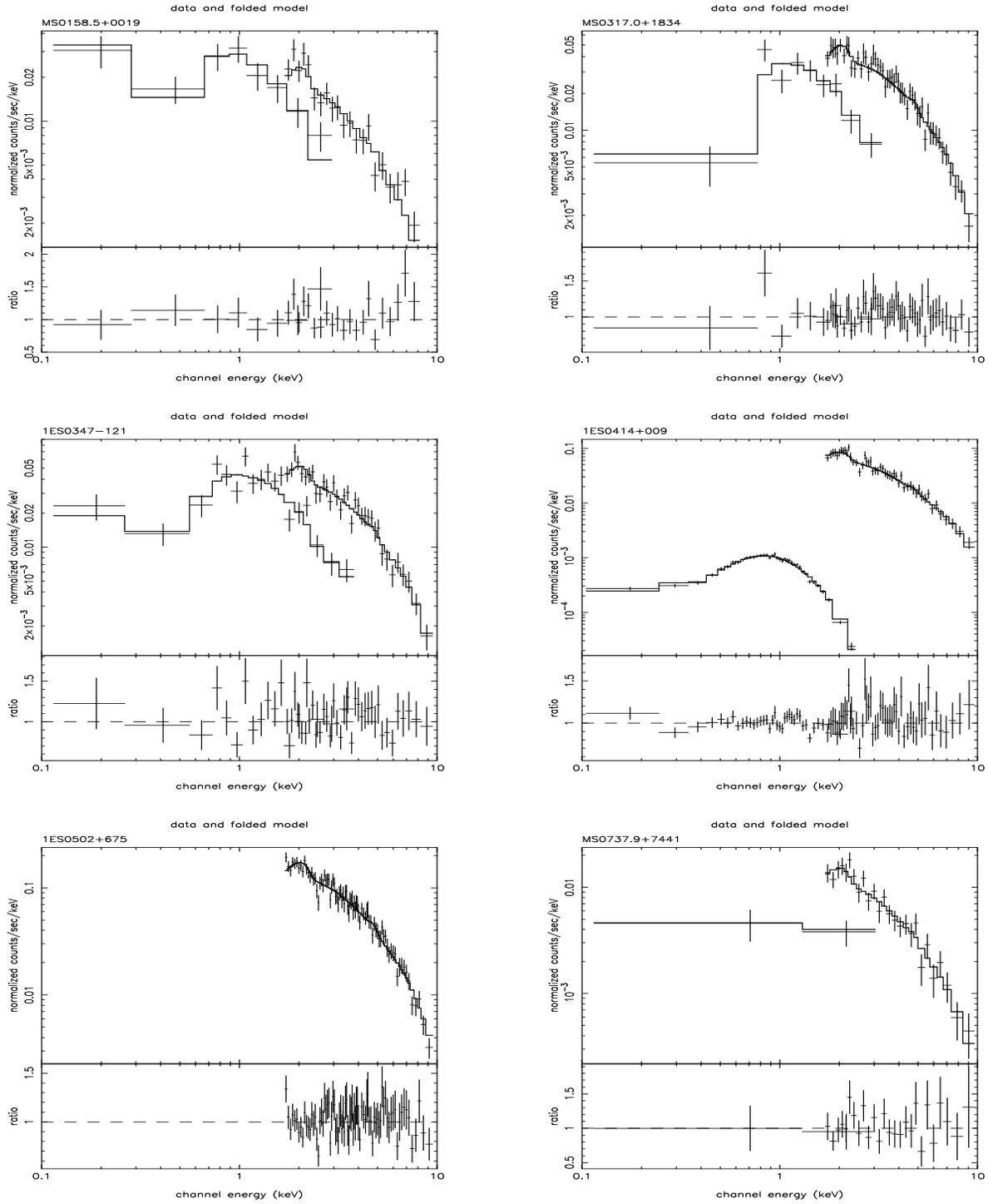}
\caption{{\bf a} X-ray (LECS and MECS) data and fitted spectrum (from Table 3 or Table 2), and ratio of data to fit.
MS0158.5+0019, MS0317.0+1834, 1ES0347--121, 1ES0414+009 (no LECS data, includes also ROSAT/PSPC data; note
that ROSAT data are in counts sec$^{-1}$ keV$^{-1}$ cm$^{-2}$), 1ES0502+675 (no LECS data),
MS0737.9+7441.} 
\label{Spec1} 
\end{figure*}

\addtocounter{figure}{-1}
\begin{figure*}[t]
\psfig{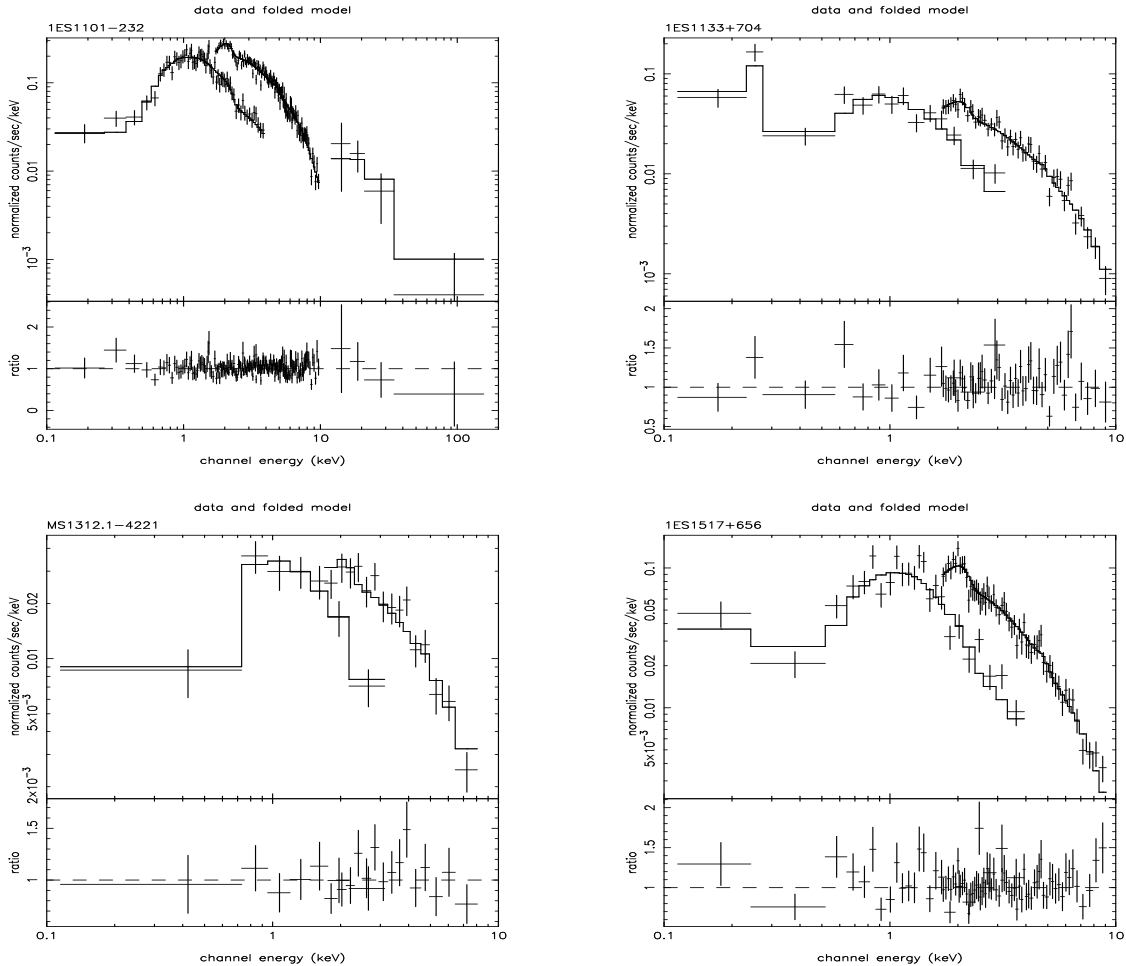}
\caption{{\bf b} (Continued) X-ray (LECS and MECS) data and fitted spectrum (from Table 3 or
Table 2), and ratio of data to fit.
1ES1101--232 (includes also PDS data), 1ES1133+704, MS1312.1--4221, 1ES1517+656.} 
\label{Spec2}
\end{figure*}

The three MECS units spectra have been summed together
to increase the S/N, after having checked that fits on
separate spectra yielded consistent results within the
statistical uncertainties.

All sources appear pointlike in both the LECS and MECS images.
Counts are extracted in a circular region of 8.5'/4' (LECS/MECS) 
radius whenever possible\footnote{i.e., when 
there is no other source in the 8.5'/4' radius; this is the case
for all sources except MS0737.9+7441, in which counts are extracted
from a region corresponding to a circle from which the bottom right 
quadrant is excluded.}. This region contains at least
95\% of the photons from the source.

The background is taken from the distributed blank sky images,
in a region corresponding to the one used to extract
source counts.
The variability of the MECS background across the field of view due to
vignetting, strongback obscuration and the spilling of 5.9 keV calibration
source photons is not known yet with high accuracy.
Given that this variability effect is higher than the secular modulation of 
the total instrumental+cosmic background, which has been estimated to be 
$\simeq 30\%$, blank fields background has been favored over an estimate
from the same image in a different, albeit close, location.
Since the background flux is never higher than 10\% of the 
source's flux, the secular modulation has a negligible
effect on the spectral results.

Extraction of the data is done in \verb+FTOOLS+ v4.0 and counts are binned so 
as to have at least 30 total counts in each bin to ensure applicability of
the $\chi^2$ statistics; the spectral analysis is performed in 
\verb+XSPEC+ v.9.0 (Shafer et al. 1991), using the matrices produced
in September 97
that include the most recent updates in calibrations. 
Net counts and errors for each source are listed in  Table~\ref{Jou}.

\medskip
We fitted together LECS and MECS, leaving free the LECS normalization 
with respect to the MECS normalization to account for the residual 
errors in intensity cross-calibration (see Cusumano, Mineo, 
Massaro et al. 1998, in preparation); the assumed spectral shape is a 
single power-law model plus free low energy absorption, arising from
cold material with solar abundances (Morrison and McCammon, 1983). 
When no LECS data are available, only the MECS data are fitted; in this
latter case the $\NH$ value is fixed to the Galactic one, since the energy range 
of the MECS does not allow a firm determination of the absorbing
column.  
Fits to the LECS data are performed only up to 4 keV, as the
response matrix of LECS is not well calibrated above this
energy (see Orr et al. 1998).

The parameters of the best fit with free $\NH$ and single power law
are given in Table~\ref{Fit} for all sources. 
Galactic $\NH$ values, as derived from the 21 cm radio survey by
Dickey and Lockman (1990) (or by the pointed observation by 
Elvis et al. 1989 when available) at the position of the source, 
are listed for reference.
Errors on the fitted parameters are
given with 90\%
confidence for one interesting parameter ($\Delta\chi^2=2.706$).
Unabsorbed fluxes (i.e., corrected for Galactic absorption)  
derived from the fit are given in the 2-10 keV band.
We give also the best fit value for the normalization factor 
of the LECS relative to MECS. The values fall in the range
expected given the current knowledge of the cross-calibration
(F. Fiore, private communication; see also 
\small\verb+http://www.sdc.asi.it/software/cookbook/cross_cal.html+).

\normalsize

\begin{table*}
\caption{Fit results for a single power law spectrum and free $N_{\rm H}$ }
\label{Fit}
\begin{tabular}[h]{| l r l l r r c r r|}
            \hline
Name   &                Energy Index& $N_{\rm H}^{\rm Gal}$ &$N_{\rm H}$& F$_{MECS}^{\rm a}$ & F$_{1{\rm keV}}$ & Norm. & $\chi^2_{\nu}$ (dof)& Prob.  \\
        &                        &    $\e20$ &    $\e20$ &(2-10)keV     & $\mu$Jy  & (LECS/MECS)   &   &   \\
            \hline
MS0158.5+0019 & $1.27_{-0.18}^{+0.18}$  & 2.67    &$ 2.9_{-1.0}^{+1.3}$     & 2.61 & 1.03 & 0.89  & 0.99 (25) & 45\%  \\ 
MS0317.0+1834 & $1.08_{-0.10}^{+0.11}$  & 10.5    &$22.0{_{-9.3}^{+13.0}}^e$  & 6.94 & 2.05 & 0.72  & 0.73 (56) & 95\%  \\ 
1ES0347--121  & $1.17_{-0.10}^{+0.11}$  & 3.64    &$ 4.7_{-1.1}^{+2.0}$     & 6.19 & 2.05 & 0.71  & 1.17 (49) & 20\%  \\ 
1ES0414+009   & $1.54_{-0.10}^{+0.10}$  & 9.15$^b$&--$^{\rm c}$             & 8.25 & 5.22 &  --   & 0.82 (45) & 80\%  \\ 
1ES0502+675   & $1.34_{-0.06}^{+0.06}$  & 9.27    &--$^{\rm c}$             &19.05 & 8.51 &  --   & 1.00 (79) & 50\%  \\ 
MS0737.9+7441$^d$&$1.53_{-0.23}^{+0.28}$& 3.54    &$25.8{_{-21.6}^{+49.3}}^e$ & 1.54 & 0.88 & 0.68  & 0.81 (24) & 72\%  \\ 
1ES1101--232  & $1.03_{-0.04}^{+0.05}$  & 5.76    &$8.9{_{-1.7}^{+3.7}}^e$    &37.92 &10.23 & 0.70  & 1.13(181) & 10\%  \\ 
1ES1133+704   & $1.47_{-0.09}^{+0.09}$  & 1.27$^b$&$3.0{_{-0.6}^{+0.8}}^f$    & 5.10 & 2.61 & 0.68  & 1.01 (62) & 48\%  \\ 
MS1312.1--4221& $1.21_{-0.19}^{+0.20}$  & 8.19    &$11.7_{-5.6}^{+11.5}$    & 4.26 & 1.52 & 0.85  & 0.74 (18) & 75\%  \\ 
1ES1517+656   & $1.44_{-0.09}^{+0.09}$  & 2.12    &$7.6{_{-1.6}^{+3.2}}^f$    &10.27 & 4.99 & 0.68  & 1.18 (79) & 15\%  \\ 
            \hline
\end{tabular}
\begin{list}{}{}
\item[$^{\rm a}$] Unabsorbed flux in $10^{-12} \ecs$
\item[$^{\rm b}$] From Elvis et al. 1989
\item[$^{\rm c}$] Fixed $\NH$
\item[$^{\rm d}$] MS0737.9+7441: There are other sources nearby. LECS count are therefore
extracted from an area 3/4 of a circle, excluding the other sources; the normalization
has been corrected multiplying by 4/3. The effect on the slope is minimal.
\item[$^{\rm e}$] marginally consistent with Galactic $\NH$ (at $\leq 2 \sigma$)
\item[$^{\rm f}$] not consistent with Galactic $\NH$ (at $\sim 3 \sigma $)
\end{list}
   \end{table*}

All the objects except 1ES1101--232 are well fitted by a power law with low 
energy absorption at a confidence level $\geq 90\%$; 
the slopes range between 1 and 1.5, with a roughly flat distribution of occurrences
and an average $\langle \alpha_x \rangle = 1.31\pm0.06$.
In two cases (1ES1133+704 and 1ES1517+656) the fitted $\NH$ 
is higher than the Galactic value by more than a factor of two (at 2.9 and 3.4
$\sigma$, respectively). 
This might indicate an intrinsic absorption, a more complex spectrum 
like a broken power law, or other features at low energy.
Furthermore, residuals are skewed at high energies in MS0158.5+0019, 
and at low energies in 1ES0347--121, 1ES1101--232 and 1ES1517+656,
even when the fit is formally acceptable.
This could be attributed to low energy absorption edges or changes 
of slope, although a residual contamination in the calibration of the
matrix at the present status is still possible (at the $\sim 15\%$
level in the 0.4-0.5 keV interval, 
Orr et al., 1998).

\noindent
To better understand the situation 
we have tried, for the eight objects for which we have the LECS data,
a more complex model for the fit, under the assumption that the
low energy absorption is not intrinsic to the object but only due to the
intervening (Galactic) material. We therefore use a broken power law with 
low energy absorption fixed to the Galactic value (as in Table~\ref{Fit})
and $\alpha_1$, $\alpha_2$ (the energy indices describing the power law 
spectral shape below and above the break energy E$_0$) and  E$_0$ free
to vary. We fix also the LECS/MECS normalization to the
values found in Table~\ref{Fit}.
In six out of eight cases an improvement (F-test probability $\geq 99 \%$)
is found over the case of fixed Galactic $\NH$ and single power
law. In the other two cases either the broken power law reduces to a single
one ($\alpha_1$ = $\alpha_2$; MS0158.5+001), or the improvement has a
marginal probability (58\%) and the parameters are poorly constrained
(MS1312.1-422).

Results of the six fits at P$\geq 99 \%$ in the broken power law case are 
listed in Table~\ref{broken}, together with the $\chi^2_{\nu}$ value
and its probability. The derived probabilities are roughly equal or
better than the corresponding probabilities found in Table~\ref{Fit}
(note that the number of free parameters in Table~\ref{Fit} is the same 
as in Table~\ref{broken} since also the normalization LECS/MECS was 
left free). This implies that the broken
power law with Galactic absorption is a good representation of the
observed spectra, from a statistical point of view.

\begin{table*}
 \caption{Fit results for a broken power law spectrum}
 \label{broken}
\begin{tabular}[h]{| l r r c r r r |}
            \hline
Name   &       Energy Index($\alpha_1$) & Energy Index($\alpha_2$) & E$_0$ & F$_{MECS}^{\rm a}$ &$\chi^2_{\nu}$ (dof) & Prob.  \\
       &                                &                          & keV   & (2-10)keV  & &    \\
            \hline
MS0317.0+1834 & 0.39 uncertain	     & 1.01(0.93-1.10)      & 0.93 uncertain  & 6.90  & 0.75(56)  & 92\% \\  
1ES0347--121  & 0.87(0.78-1.08)      & 1.23(1.13-1.33)      & 1.45 uncertain  & 6.10  & 1.10(49)  & 29\% \\
MS0737.9+7441 & 0.17 uncertain       & 1.43(1.27-1.61)      & 1.18(0.6 - 3.)  & 1.51  & 0.82(24)  & 71\% \\
1ES1101--232  & 0.59(0.35-0.74)      & 1.05(1.01-1.08)      & 1.36(1.11-1.65) &37.63  & 1.06(181) & 28\% \\
1ES1133+704   & 0.85($<$1.13)        & 1.47(1.39-1.57)      & 0.93($<$2.13)   & 5.08  & 1.01(62)  & 45\% \\
1ES1517+656   & 0.40(0.15-0.59)      & 1.48(1.41-1.56)      & 1.25(1.08-1.43) &10.09  & 1.02(79)  & 43\% \\
            \hline
\end{tabular}
\begin{list}{}{}
\item[$^{\rm a}$] Unabsorbed flux in $10^{-12} \ecs$
\end{list}
\end{table*}

A good determination of the confidence interval (90\% level for
one interesting parameter) for all the parameters
is possible only for the two sources with higher statistics, however the 
$\alpha_2$ values are generally well constrained and consistent with the slope 
of the single power law fit, while the $\Delta\alpha$ is on average about 0.5, 
consistent with other results for BL Lacs
(e.g., Sambruna et al. 1996; Urry et al. 1996). Therefore, the hypothesis that
the apparent excess in absorption could instead be due to an intrinsically 
curved (convex) spectrum seems to be plausible.

The brightest source of the sample (1ES1101--232) has been detected 
(at 3.5$\sigma$) also in the PDS  in just $\sim$ 6ks of on-source observing time. 
We have therefore a measurement at the same epoch of the spectrum between
0.1 and $\sim$ 100 keV (see spectrum in Figure~\ref{Spec1}).
Net on-source PDS spectra have been obtained simply subtracting ``off-~'' from
``on-position'' ones.
The spectra of the four units have then been summed together after energy
equalization.
 Very short
($\tau < 1 \ s$) and intense spikes, due to single-particle-induced
fluorescence in the crystals, have been removed applying the method
described in Matt et al. (1997).
We fitted the PDS spectrum together with LECS and MECS spectra: the slope is 
flat and consistent with the same slope as the MECS above $\sim 1.5$ keV 
(see $\alpha_2$ value from Table~\ref{broken}). 
This results imply that the harder energy band  is within
reach with exposures of some tens of ksec for many objects of the sample,
especially those with a flat spectrum.

Spectra for the 10 sources are shown in Figure~\ref{Spec1}, including
the best fit model derived from the broken power law model for the
six sources of Table~\ref{broken}, and from the single power law model 
for the other four sources, and including the PDS data for 1ES1101--232, 
and the ROSAT/PSPC data for 1ES0414+009 (see Section 4).

We have also computed the upper limits in the PDS, under the assumption that the
spectral slope derived in the MECS continues into the PDS band, and plotted 
the results in the overall
spectra presented in Figure~\ref{FigSeda} (see Section 6).
No source was detected in the HPGSPC, which has a better energy
resolution but lower effective area than the PDS.

\section{Comparison with ROSAT Observations}

\subsection{Data and spectral Analysis}

Eight out of ten sources have been observed by the PSPC instrument
onboard ROSAT and some of them have been  
previously published by different authors (the EMSS
sources, Perlman et al. 1996b). In order to ensure
a uniform procedure for all the sample we have re--analyzed 
them, obtaining results consistent with those published.
We have processed the data with a standard reduction procedure, as
described e.g. in Comastri Molendi \& Ghisellini (1995).

\begin{table*}
\caption[]{Fits in the 0.1--2 keV energy range, using ROSAT data.}

\noindent
{\hfill\begin{tabular}{|l l l c l l c l|}
\hline
Name & $\NH$ & $F^a_{1~\rm keV}$ & $\alpha_{ROSAT}$ &  $\chi^2_{\nu}$(d.o.f.) & $N_{\rm H Gal}$ &
 $\alpha_{ROSAT}$ &  $\chi^2_{\nu}$(d.o.f.) \\
      & $\e20$    &  $\mu$Jy  &    &  & $\e20$ &  &     \\
\hline
MS0158.5+0019 & 3.0(2.3--3.6) & 1.34 & 1.45(1.23--1.68) & 0.63 (14) & 
   2.67 & 1.35(1.28--1.43) & 0.63 (15) \\ 
MS0317.0+1834 & 9.5(6.7-14.1) & 0.56 & 1.50(1.02--2.04) & 1.81 (13) & 
   10.5 & 1.63(1.32--1.91) & 1.70 (14) \\ 
1ES0347$-$121 & 3.6(3.3--3.8) & 2.55 & 1.10(1.02--1.19) & 1.01 (42) & 
   3.64 & 1.13(1.09--1.16) & 1.00 (43) \\ 
1ES0414+009 & 9.5(9.1--10.0) & 4.69 & 1.63(1.56--1.71) & 1.20 (36) & 
   9.15 & 1.58(1.54--1.62) & 1.22 (37)\\ 
MS0737.9+7441 & 4.1(3.8--4.5) & 1.73 & 1.34(1.23--1.45) & 1.38 (37) & 
   3.54 & 1.16(1.13--1.20) & 1.54 (38) \\ 
1ES1101$-$232 & 6.8(6.4--7.1) & 11.2 & 1.43(1.35--1.51) & 1.08 (35) & 
   5.76 &  1.23(1.20--1.27) & 1.67 (36) \\ 
1ES1133+704 &  1.3(1.1--1.6) & 2.60 & 1.31(1.22--1.40) & 1.94 (32) & 
   1.27 & 1.29(1.26--1.32) & 1.89 (33) \\ 
1ES1517+656 &  3.6(3.3--3.9) & 7.42 & 1.29(1.19--1.39) & 1.14 (49) & 
   2.12 & 0.82(0.78--0.85) & 2.49 (50) \\ 
\hline
\end{tabular}\hfill}

$^a$ Model flux at 1~keV.  \\
\label{TabRosat}
\end{table*}

All the 8 sources have been clearly detected with
enough counts to allow spectral analysis.
The background--subtracted light curves are not variable over the 
entire observation, typically lasting 3--9 ksec with maximum deviations
of the order of 30--40 \%, likely due entirely to instrumental effects.
In principle however variations on time scales of the order of $\sim 0.5$ 
ksec were accessible.
The source spectra were rebinned in order to obtain a significant
S/N ($>5$) for each bin and fitted with a single power--law model 
plus absorption arising from cold material with solar
abundances (Morrison \& McCammon 1983).
The derived spectral parameters are given in Table~\ref{TabRosat}, where the reported
errors are at 90\% confidence level
for one interesting parameter. 
All the spectra were fitted with  the column density i)
fixed at the Galactic value, and ii) free to vary. 
Values for the Galactic column densities towards the objects in our sample 
are reported for ease of reference from Table~\ref{Fit}.
\par

In most cases a single power law spectrum with the absorption fixed at 
the Galactic value provides an excellent description of the data, 
while for some of the objects either the fits are statistically
unacceptable and/or the column density inferred from the fit
is not consistent with the Galactic value, 
thus requiring a more complex description of the spectral shape.

\subsection{Comparison between \sax\ and ROSAT Results}

\begin{figure}
\psfig{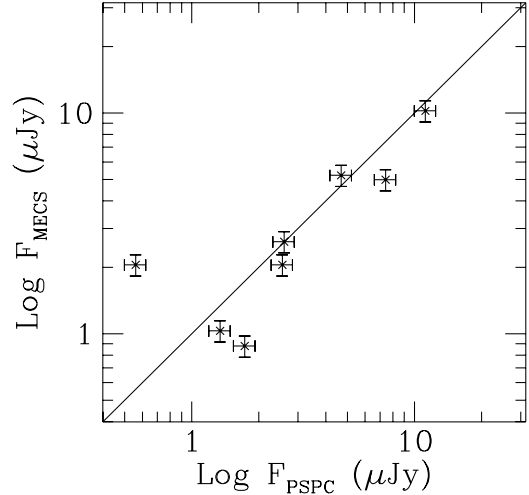}
\caption{X-ray flux at 1 keV from \sax\ vs. X-ray flux at 1 keV from
ROSAT in $\mu$Jy. Errors are 10\% of the flux.
The solid line represents the locus of $F_{MECS}=F_{PSPC}$.}
\label{FigFlux}
\end{figure}

\begin{figure}
\psfig{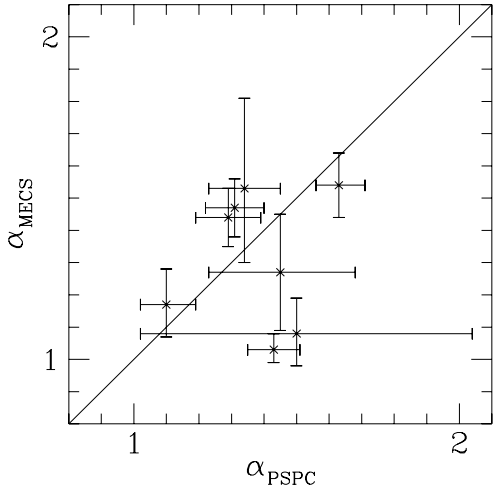}
\caption{$\alpha_x$ from \sax\ vs. $\alpha_{PSPC}$ from
ROSAT. The solid line represents the locus of $\alpha_x$=$\alpha_{PSPC}$.}
\label{FigAlpha}
\end{figure}

We compare the \sax\ and ROSAT data as taken from Table~\ref{Fit}
and Table~\ref{TabRosat},
 respectively. We plot the two fluxes at
1 keV in Figure~\ref{FigFlux}, assuming total errors to be
10\% of the flux (higher than those from counting statistics only
and taking into account residual absolute calibrations);
the solid line represents the $F_{MECS}=F_{PSPC}$ locus. From 
the comparison of the ROSAT and \sax\ results it emerges
that, although these observations cover a time interval of few years,
the 1 keV fluxes of the various sources are practically the same within
30\% for all but three sources: MS0317.0+1834, with a \sax\ 
flux $\sim 4 \times$ larger, and
MS0737.9+7441, and 1ES1517+656 with a ROSAT flux $\sim 2$ and $\sim 1.5 \times$
larger, respectively. 

The ROSAT and \sax\ single power law spectral indices, plotted
one against the other in Figure~\ref{FigAlpha}, are also consistent
for all but one source: 1ES1101--232; in particular the PSPC
spectrum of this source can be fitted also by a broken power law,
with $\NH$ fixed at the Galactic value\footnote{Results of the fit: 
$\alpha_1$ = 1.07 (0.85 - 1.16); $\alpha_2$ = 1.42 (1.33-1.54); 
E$_0$ = 0.69 (0.44-0.93) keV with a $\chi^2$ = 42.3 (33 dof) and 
an F-test probability with respect to the fixed Galactic $\NH$ and single power law of 99.99\%.} 
but the values of the parameters are inconsistent with those found in the \sax\
observation.
This may be an indication that, even if the source flux is
roughly the same, in 1ES1101--232 there has been a variation of the spectrum
and in particular of the position of the peak of the synchrotron
emission.

Instead, with respect to the $\NH$ values as determined with the two
satellites, the situation is more complex. First of all, the ROSAT values
are probably more reliable than the \sax\ ones, given the small number of
counts usually detected in the LECS for our sources (only for 1ES1101--232
we have good statistics in the LECS). The ROSAT $\NH$ values are consistent
with the Galactic ones for five sources out of eight while for the other
three (MS0737.9+7441, 1ES1101--232 and 1ES1517+656) the F-test shows that
introducing $\NH$ as a free parameters improves the $\chi^2$ significantly
(97.4\%, 99.99\%, and $>$99.99\%, respectively). 
For  MS0158.5+0019 and 1ES0347-121 the $\NH$ measured by the PSPC is
also consistent with that measured from the LECS/MECS data. In the other cases
the PSPC sees a lower amount of $\NH$ (and with smaller error intervals)
than the LECS/MECS data. 

Again these are clear indications in favor of a more
complex spectrum than a simple power law (see section 3.2). Indeed, a broken
power law with $\NH$ fixed to the Galactic value seems a better representation
to the \sax\ data for all these five sources (see Table~\ref{broken}).

For 1ES0414+009, for which the intensities in the MECS and
PSPC are comparable, and for which we lack a determination
of the low energy portion of the spectrum (no LECS data),
we combine the two instruments, PSPC and MECS for a simultaneous
fit (see Figure~\ref{Spec1}). The slope is $\alpha = 1.59\pm0.6$,
consistent with $\alpha_x$ from the MECS only, and $\NH$ = 9.3$\pm0.4 
\times 10^{20}$ cm$^{-2}$, consistent with the Galactic value.

\section{Same--Epoch Optical Observations}

In the course of an optical observing campaign performed at the Torino
Observatory, the eight sources with declination greater than $-20^\circ$ were
observed during or close to the time of \sax\ pointing.
The data were taken with the 1.05 meter REOSC telescope.
The
equipment includes a $1242 \times 1152$ pixel CCD detector with a 0.47 arcsec
per pixel scale and standard Johnson's $BV$ and Cousins' $R$ filters. 

Data reduction was obtained with the Robin procedure developed at
the Torino Astronomical Observatory (Villata et al.\ 1997), which
performs bias subtraction, flat field correction, and circular Gaussian fit
after background subtraction. 

Magnitude calibration was derived through comparison with reference stars in
the same field of the source; for MS0158.5+0019 and MS0317.0+1834 we used the
photometric sequences published by Smith et al.\ (1991), while for the other BL
Lacs we adopted our own sequences: those in the field of 1ES0502+675, MS
0737.9+7441, 1ES1133+704, and 1ES1517+656 are published in Villata et al.\
(1998). 

Optical light curves of the eight BL Lacs in the period around the \sax\
pointings are shown in Raiteri et al.\ (1998). 
Table~\ref{Mag} gives the data closest in time to the SAX observations.

\begin{table*}
 \caption{Same-epoch Optical Magnitudes}
  \label{Mag}
\begin{tabular}[h]{| l r c c c c|}
\hline
  SOURCE  &   \sax\ date     & OPTICAL date &    R      &     V     &      B     \\
\hline
MS0158.5+0019 & 16-17 Aug 96 & 22 Oct 96 &            & 18.60$\pm$.10 &  	\\
              &              & 31 Oct 96 &  $>$17.7   &            &  	\\
\hline
MS0317.0+1834 &   15 Jan 97  & 18 Dec 96 &  18.10$\pm$.08&            &   	\\
              &              & 16 Jan 97 &  17.83$\pm$.08&            &   	\\
              &              & 16 Jan 97 &            & 18.33$\pm$.10 &  	\\
              &              & 29 Jan 97 &  18.02$\pm$.08&            &   	\\
\hline
1ES0347-121 &   10 Jan 97    & 17 Dec 96 &  17.38$\pm$.08&            &   	\\
              &              & 10 Jan 97 &  17.37$\pm$.08&            &   	\\
              &              & 13 Jan 97 &  17.33$\pm$.08&            &   	\\
              &              & 13 Jan 97 &            & 18.10$\pm$.10 &  	\\
\hline
1ES0414+009 &21-22 Sep 96    & 19 Oct 96 &            & 16.86$\pm$.04 &  	\\
              &              & 23 Oct 96 &            & 16.86$\pm$.04 &  	\\
\hline
1ES0502+675 &06-07 Oct 96    & 18 Oct 96 &            & 17.30$\pm$.04 &  	\\
              &              & 23 Oct 96 &            & 17.01$\pm$.04 &   	\\
\hline
MS0737.9+7441 &29-30 Oct 96  & 07 Nov 96 &  17.44$\pm$.08&            &   	\\
              &              & 27 Nov 96 &  17.29$\pm$.08&            &   	\\
\hline
1ES1133+704 &10-11 Dec 96    & 18 Dec 96 &  14.85$\pm$.04&            &   	\\
              &              & 14 Jan 97 &  14.94$\pm$.04&            &   	\\
              &              & 14 Jan 97 &            & 15.38$\pm$.04 &  	\\
              &              & 14 Jan 97 &            &            & 16.04$\pm$.13 \\
\hline
1ES1517+656 &   05 Mar 97    & 02 Mar 97 &  16.11$\pm$.03&            &   	\\
              &              & 05 Mar 97 &  16.11$\pm$.03&            &   	\\
              &              & 06 Mar 97 &  16.08$\pm$.03&            &   	\\
              &              & 10 Mar 97 &  16.08$\pm$.03&            &   	\\
              &              & 13 Mar 97 &  16.08$\pm$.03&            &   	\\
              &              & 13 Mar 97 &            & 16.43$\pm$.03 &  	\\
              &              & 13 Mar 97 &            &            & 16.78$\pm$.05 \\
\hline
\end{tabular}
\end{table*}

Using the same epoch optical magnitudes for the 8 objects north
of $-20^\circ$, and literature magnitudes for 1ES1101--232 and
MS1312.1--4221, we compute the two point spectral indices
$\alpha_{ox}$ and $\alpha_{ro}$, reported in Table~\ref{alpha}
together with $\alpha_{rx}$, log($F_x/F_r$)and  $L_x$, and 
$\nu_{peak}$ computed in the next Section.
The frequencies used to compute the broad band indices are 5 GHz, 2500 \AA\,
and 2 keV. 
We apply a K--correction using $z=0.2$ for the two sources with unknown redshift, 
$\alpha_r =0.2$, $\alpha_o=0.65$ and $\alpha_x$ from Table~\ref{Fit}. 

\begin{table*}
\caption{Derived quantities: Two-point overall spectral indices ($\alpha_{ox}$;
$\alpha_{ro}$; $\alpha_{rx}$), X-ray--to--radio flux ratio, logarithm
of X-ray luminosity, and logarithm of $\nu_{peak}$.}
\label{alpha}
\begin{tabular}[h]{| l r r r r l r |}
\hline
Name  &      $\alpha_{ox}$ & $\alpha_{ro}$ & $\alpha_{rx}$ & $\log(F_x/F_r$) & $\log L_x$ & $\log \nu_{peak}$ \\
\hline
MS0158.5+0019   &   0.89 &  0.44 &  0.59  &  -9.71 &  45.12 &  16.73 \\
MS0317.0+1834   &   0.78 &  0.44 &  0.55  &  -9.39 &  45.11 &  17.01 \\
1ES0347$-$121   &   0.83 &  0.37 &  0.52  &  -9.16 &  45.05 &  16.97 \\
1ES0414+009     &   0.95 &  0.49 &  0.64  & -10.16 &  45.59 &  16.47 \\
1ES0502+675     &   0.76 &  0.42 &  0.53  &  -9.24 &  45.6 $\rm^a$&  16.83 \\ 
MS0737.9+7441   &   1.09 &  0.43 &  0.64  & -10.19 &  44.93 &  15.97 \\
1ES1101$-$232   &   0.76 &  0.47 &  0.57  &  -9.50 &  45.83 &  17.48 \\
1ES1133+704     &   1.29 &  0.44 &  0.71  & -10.64 &  43.69 &  15.65 \\
MS1312.1$-$4221 &   1.12 &  0.32 &  0.58  &  -9.63 &  44.85 &  15.90 \\
1ES1517+656     &   0.99 &  0.37 &  0.57  &  -9.56 &  45.3 $\rm^a$&  16.60 \\
\hline
\end{tabular}
\begin{list}{}{}
\item[$\rm^a$] Redshift unknown; z=0.2 assumed to compute luminosity
\end{list}
\end{table*}

\begin{figure*}
\psfig{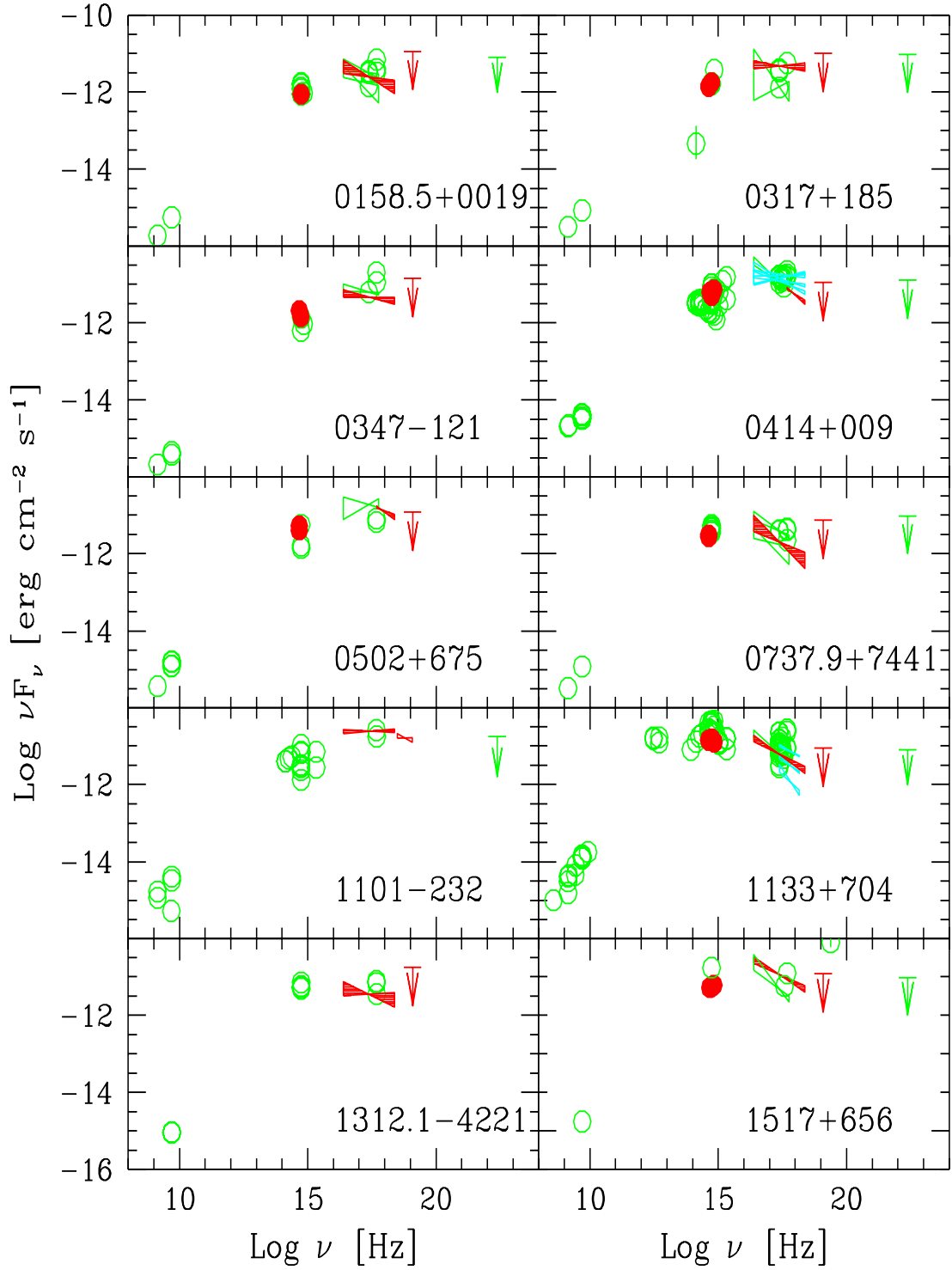}
\caption[h]{The spectral energy distribution of the XBLs.
Data from \sax\ spectra and same epoch optical magnitudes from this
paper are in filled symbols, data from literature (open symbols or
upper limits) are referenced in Table~\ref{tab:ref}.
The name of the object is indicated in each panel.}
\label{FigSeda}
\end{figure*}

\begin{table*}
\caption[]{ References for data shown in Fig.~\ref{FigSeda}
}
\begin{tabular}{l l} 
& \\
\hline
& \\
MS0158.5$+$0019 &G95, J94, L96, P96a, P96b, N96, W94 \\
MS0317.0$+$1834 &G93, G95, L96 \\
1ES0347$-$121   &G95, L96, P96a    \\
1ES0414$+$009   &B92, F93, G95, GC91, Gr95, IT88, L96, M90, P96a, Pi94, Pi93, 
                Sa94  \\
1ES0502$+$675   &B94, B95, G95, GC91, P96a  \\
MS0737.9$+$7441 &G95, J94, L96, P96a, PS93, W94 \\
1ES1101$-$232   &B92, E92, F93, F94, G95, L95, P94, P96a, Pi93, R89  \\
1ES1133$+$704   &B95, Bi92, C95, E92, G90, G95, GS95, IN88, K91, L96, MB86, \\
                &N96, P92, P96a, PG95, Pi93, Sa94, T84, WW90 \\
MS1312.1$-$4221 &G95, P96a, S90  \\
1ES1517$+$656   &B94, G95, MB95, P96a  \\
& \\
\hline
& \\
\end{tabular}
References:
B92: Bersanelli et al., 1992;
B94: Brinkmann et al., 1994;
B95: Brinkmann et al., 1995;
Bi92: Biermann et al., 1992; 
C95: Ciliegi et al., 1995;
E92: Edelson et al., 1992;
F93: Falomo et al., 1993;
F94: Falomo et al., 1994;
G90: Giommi et al., 1990;
G93: Gear, 1993;
G95: Giommi et al., 1995;
GC91: Gregory \& Condon, 1991;
Gr95: Griffith et al., 1995;
GS95: Ghosh \& Soundararajperumal, 1995;
IN88: Impey \& Neugebauer, 1988;
IT88: Impey \& Tapia, 1988;
J94: Jannuzi et al., 1993;
K91: Kinney et al., 1991;
L95: Lanzetta et al., 1995 ;
L96: Lamer et al., 1996;
M90: Mead et al.,  1990;
MB86: Mazzarella \& Balzano, 1986;
MB95:  McNaron--Brown et al., 1995;
N96: Nass et al., 1996;
P92: Patnaik et al., 1992;
P94: Pesce et al., 1994;
P96a: Perlman et al., 1996a;
P96b: Perlman et al., 1996b;
PG95: Padovani \& Giommi, 1995;
Pi93: Pian \& Treves, 1993;
Pi94: Pian et al., 1994;
PS93: Perlman \& Stocke, 1993;
R89: Remillard et al., 1989;
S90: Stocke et al., 1990;
Sa94: Sambruna et al., 1994;
T84: Tovmassian et al., 1984;
W94: Wolter et al., 1994;
WW90: Worrall \& Wilkes, 1990.
\label{tab:ref}
\end{table*}

\section{Spectral Energy Distributions}

In Fig. \ref{FigSeda} we plot the spectral energy distribution
(SED) of our sources, reporting the \sax\ spectral fits and the nearly
simultaneous optical data when available (filled symbols), and other data
from the literature (open symbols or upper limits; references 
are listed in  Table~\ref{tab:ref}).

As can be seen, the SED of the objects in our sample are 
characterized by a smooth distribution, rising (in $\nu F_\nu$) 
from the radio to the optical-UV and X-ray bands. 
Variability is evident, especially in the most often observed objects. 
Many objects show a fall off at high energies.
In order to obtain a reliable estimate of the peak energy, the SED of each
object has been fitted with a polynomial function of the type
\begin{equation} 
\log(\nu F_\nu ) = a + b (\log \nu) + c (\log \nu)^2 + d (\log \nu)^3
\end{equation}
(see Comastri et al. 1995).
For the fit, we adopted the \sax\ data for the X-ray band, and same-epoch 
optical data when available. 
For the other bands (i.e., radio and IR --  none of these sources has 
been detected in  $\gamma$-rays) we used the maximum observed value:
variability amplitudes are not large in these bands and so 
should not affect significantly the results. From the fitted polynomial 
we derive the peak frequency of the SED $\nu_{peak}$.

Three of the sources in this sample have been studied also by Sambruna 
et al. (1996), where the $\nu_{peak}$ value was derived assuming a 
parabolic fit.
If we compare the $\nu_{peak}$ values derived here and
in Sambruna et al. (1996), in two cases (MS0158.5+0019 and MS0737.9+7441) 
we have consistent results, while in the third (MS0317.0+1834) $\nu_{peak}$ 
differs by about 2.5 orders of magnitudes ($\log \nu_{peak}=17.01$ 
vs. $\log \nu_{peak}=14.36$ in Sambruna et al. 1996).
Note, however, that their Figure 4 shows that the peak of the emission
for MS0317.0+1834 is indeed at higher energies, in agreement with our value.
Note also that the optical simultaneous observations constrain 
$\nu_{peak}$ in a much better way.

\begin{figure}
\psfig{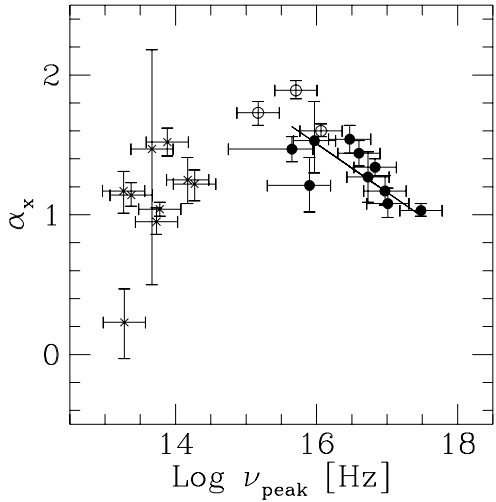} 
\caption[h]{$\alpha_x$ as a function of $\nu_{peak}$.
Filled circles are HBL with $\alpha_x$ values from \sax\ as in 
Table~\ref{Fit}, $\nu_{peak}$ from fit of SED -- see Section 6.
Empty circles (HBL) and crosses (LBL) are from Comastri et al. (1995) 
in which $\alpha_x$ is derived from ROSAT observations. 
The solid line represent the linear regression (with the OLS method)
between the 10 HBL observed by \sax.}
\label{FigAlN}
\end{figure}

Figure~\ref{FigAlN} shows the spectral indices derived
from the LECS/MECS fit (from Table~\ref{Fit}) vs. log $\nu_{peak}$. 
We include also the data for the objects studied by Comastri et al. (1995),
which are 9 LBL (crosses) and 3 HBL (empty circles).
These latter points are derived from ROSAT observation and so 
sample the spectra at slightly lower energy. 
A program is underway to observe a sample of BL Lacs extracted
from the 1 Jy sample (Padovani et al. 1998 and in preparation). 
When a sizable sample of LBL will be observed by \sax, a more direct 
comparison will be possible.
For the HBL objects 
there is a good anti-correlation between $\alpha_x$ and $\nu_{peak}$
(correlation probability = 95\%).
The solid line represents the linear regression obtained with the
10 objects observed by \sax\ using the
OLS bisector method (Isobe et al. 1990).
The same-epoch data used for most sources, that remove the uncertainty due to 
variation for the determination of $\nu_{peak}$, are important in order
to obtain such a tight correlation.
The LBL are less correlated
(correlation probability $\sim 80\%$). 

\begin{figure}
\psfig{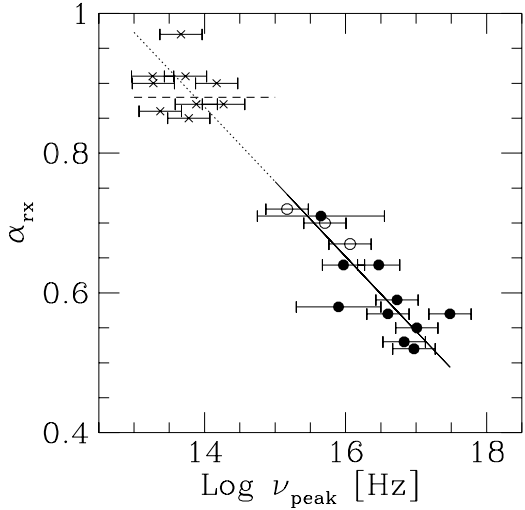} 
\caption[h]{$\alpha_{rx}$ as a function of $\nu_{peak}$.
Filled circles are HBL: $\alpha_{rx}$ values from Table~\ref{alpha}, 
$\nu_{peak}$ from fit of SED -- see Section 6.
Empty circles (HBL) and crosses (LBL) are from Comastri et al. (1995) 
in which $\alpha_x$ is derived from ROSAT observations. 
The solid line represents the linear regression fit using the OLS method on the
10 objects observed by \sax\, while
the dotted line indicates its extrapolation to lower values of $\nu_{peak}$.
The horizontal dashed line ($\alpha_{rx}=$ constant) is derived
from the model described in Padovani \& Giommi 1995 for LBL objects.}
\label{FigArx}
\end{figure}

The behavior of $\alpha_x$ with $\nu_{peak}$, illustrated in 
Figure~\ref{FigAlN}, reflects the large range of frequencies 
at which the synchrotron peak is observed to lie. 
This picture is consistent with the one proposed by Padovani and Giommi (1996) 
using ROSAT data, in which 
HBL are characterized by a synchrotron peak located
in the EUV/X--ray band: the X--ray radiation is therefore
produced by the synchrotron process.
According to the exact location of $\nu_{peak}$,
the X--ray spectrum changes, being flat ($\alpha_x \le 1$) in extreme 
HBL objects (where $\nu_{peak}$ lies in the X--ray band), and steeper
as the peak migrates to the UV band. 
In LBL objects $\nu_{peak}$ has moved in the optical-IR band and 
the X--ray band is dominated by the flat inverse Compton spectrum.

In Figure~\ref{FigArx} we plot for both these ten HBL with \sax\ data 
and the Comastri et al. (1995) objects
the relationship between $\alpha_{rx}$ and log $\nu_{peak}$.
We add the linear regression obtained using the OLS bisector method
with the \sax\ data only (solid line) and extrapolated down to the LBL region 
(dotted line), besides the constant $\alpha_{rx}$ (horizontal dashed line) derived 
following Padovani \& Giommi (1995) (see their Figure 12). 
The HBL points are
correlated at the 98\% level, indicating a link between the
two quantities, while the LBL points are consistent
both with the extrapolation of the regression, or with the constant
value for $\alpha_{rx}$ suggested by Padovani \& Giommi (1995),
in the hypothesis that for LBL the X-ray flux is proportional
to the radio flux.

Again, it will be indeed interesting to compare these results 
with those obtained for LBL with \sax\ data (Padovani et al., in preparation).

\section{Results and Conclusions}

We have analyzed the spectra for 10 X-ray selected BL Lacs observed with the
Narrow Field Instruments on board the \sax\ satellite.

The sources are detected from $\sim 0.2$ up to $\sim$ 10 keV (and in one 
case up to $\sim 100$ keV with the PDS instrument) 
with a very smooth appearance. The spectrum is generally well fitted
by either a single power law, or by a broken convex power law that
most probably represents the steepening after the synchrotron peak,
whose position is determined also by using simultaneous optical observations.

Variability is not present during the short \sax\ exposure;
analysis of ROSAT data shows for most of the sources little variability 
(within $\sim 30\%$) with respect to the \sax\ flux  and 
spectral indices consistent with the \sax\ ones. The spectral energy 
distributions, which include literature data, instead show variability in all bands. 

The X-ray spectral indices $\alpha_x$ range between 1 and 1.5 with a flat
distribution and a mean value $\langle \alpha_x \rangle = 1.31\pm0.06$.  The
scatter in the distribution is due to an anti-correlation we have found
between $\alpha_x$ and the frequency of the peak of the emission,
$\nu_{peak}$. This extends to the \sax\ band a correlation which had been
discovered in the ROSAT band for this class of objects. The fact that sources 
with
harder X-ray spectra have higher $\nu_{peak}$ is expected if the \sax\ band 
is still dominated by synchrotron emission, which is also consistent with the
spectral energy distributions of our BL Lacs.

Furthermore, we have no evidence of a spectral flattening (indicating
the arising of the Compton component) in the present spectra, but
future PDS detections, that are possible with exposure times 
slightly longer than those obtained here, might help in this respect.

The large fraction (at least 2 out of 10) of HBL
selected in the soft X-ray band found with a flat ($\alpha_x \sim 1$) 
X-ray slope (i.e., they are near the peak of the synchrotron emission)
and the distribution of $\alpha_x$ values support
the view that objects with even higher spectral peaks in their
quiescent status indeed exist,
and might be found in large numbers if we devise the correct
strategy (e.g., samples at harder X-rays, TeV sources, etc.)

Moreover, these sources are good candidates to be TeV {\it emitters}.
In fact, in the sources with the flattest $\alpha_x$ the peak of the synchrotron
component is localized in the soft X--ray range.
Electrons emitting at 1 keV by the synchrotron process have Lorentz factors
$\gamma\sim 2.5\times 10^5 (\nu_{peak,1~keV}/B\delta)^{1/2}$,
where $\nu_{peak} = 2.42 \times 10^{17} \nu_{peak,1~keV}$ Hz,
$B$ is the value of the magnetic field in Gauss and $\delta$ is
the usual Doppler factor.
Through the inverse Compton mechanism, they can emit up to
$E\sim \gamma m_ec^2\delta\sim 130 (\nu_{peak,1~keV}\delta/B)^{1/2}$ GeV.
If the magnetic and radiation energy densities are equal
(as it is, approximately, in the three BL Lacs already
detected in the TeV band), the flux level of the synchrotron and
inverse Compton peaks is roughly equal.
Low redshift sources are therefore good
candidates to be {\it detected} in the GeV--TeV band, while
the high energy emission of the more distant sources could be
absorbed in $\gamma$--$\gamma$ interactions with the background
IR photon field, whose intensity is still uncertain.
Indeed, a cutoff in the high energy spectrum could be used to
determine the IR background (see, e.g. Stecker \& De Jager, 1997).

This \sax\ project is still ongoing. We expect therefore to increase 
considerably the sample of soft X-ray selected BL Lacs for which we
measure the spectrum in the 0.2-10 keV range and possibly above.  
With a larger complete sample, and combining the results with
other complementary \sax\ projects, we expect to be able to
draw a clearer picture of the relationship between the local
X-ray slope and the overall energy distribution of this
class of sources,
in order to derive firmer conclusions on the behaviour at hard
X-ray energies and on the mechanisms of the emission.

\begin{acknowledgements}
This work has received partial financial support from the Italian 
Space Agency (ASI contracts ASI 95/RS 72/SAX, and ASI ARS-96-70 ).

\end{acknowledgements}


\begin{thebibliography}{}

\bibitem{} Bersanelli M., Bouchet P., Falomo R., Tanzi E.G., 1992, AJ, 104, 28 (B92)

\bibitem{} Biermann P.L., Schaaf, R., Pietsch, W., Schmutzler, T., Witzel, A. \& 
Kuhr, H., 1992, A\&AS 96, 339 (Bi92) 

\bibitem{} Boella G., Butler R.C., Perola G.C., et al., 
		1997a, A\&AS 122, 299

\bibitem{} Boella G., Chiappetti L., Conti G., et al., 
		1997b, A\&AS 122, 327

\bibitem{} Bregman, J.N. 1990, A\&ARev, 2, 125

\bibitem{} Brinkmann W., \& Siebert J., 1994, A\&A 285, 812 (B94) 

\bibitem{} Brinkmann W., Siebert J., Reich W., Furst E., Reich P., Voges W.,
     Trumper J., Wielebinski R., 1995, A\&AS, 109, 147 (B95)
              
\bibitem{} Butler C., Scarsi L., 1990, SPIE 1344, 46

\bibitem{} Catanese, M. et al., 1997, ApJL, 487, L143

\bibitem{} Catanese, M. et al., 1998, preprint, astro-ph/9712325

\bibitem{} Ciliegi P., Bassani L. \& Caroli E., 1995, ApJ 439, 80 (C95)

\bibitem{} Comastri A., Molendi S., and Ghisellini,G.
1995, MNRAS, 277, 297

\bibitem{} Dickey J.M. \& Lockman F.J. 1990, ARAA, 28, 215

\bibitem{} Edelson R., Pike G.F., Saken J.M., Kinney A., \& Shull, J.M.
        1992, ApJS 83, 1 (E92) 

\bibitem{} Elvis M., Wilkes, B.J., \& Lockman, F.J., 1989, AJ, 97, 777

\bibitem{} Falomo R., Bersanelli M., Bouchet P., Tanzi E.G., 1993, AJ, 106, 11 (F93) 

\bibitem{} Falomo R., Scarpa R., Bersanelli M., 1994, ApJS, 93, 125 (F94)

\bibitem{} Gear W.K., 1993, MNRAS 264, 919 (G93) 

\bibitem{} Ghosh K.K. \& Soundararajaperumal S., 1995, ApJS 100, 37 (GS95) 

\bibitem{} Giommi P., Barr P., Garilli B., Maccagni D. \& Pollack A.M.T., 1990, 
        ApJ 356, 432 (G90) 

\bibitem{} Giommi P., Ansari S.G., Micol A., 1995, A\&AS, 109, 267 (G95) 

\bibitem{} Giommi P., Padovani P. \& Perlman, E., 1998,
To appear in ``The Active X-ray Sky: Results from BeppoSAX and Rossi-XTE'', 
Nuclear Physics B Proceedings Supplements, L. Scarsi, H. Bradt, P. Giommi 
\& F. Fiore (eds.), Elsevier Science B.V

\bibitem{} Gregory P.C. \& Condon J.J., 1991, ApJS 75, 1011 (GC91) 

\bibitem{} Griffith M.R., Wright A.E., Burke B.F. \& Ekers R.D., 1995, ApJS, 97, 347 (Gr95) 

\bibitem{} Guainazzi M. \& Matteuzzi A., 1997, SDC-TR 011

\bibitem{} Hasinger G., Turner T.J., George I.M. \& Boese G., 1992, Legacy, 2

\bibitem{} Impey C.D., Neugebauer G., 1988, AJ, 95, 307 (IN88)

\bibitem{} Impey C.D., Tapia S., 1988, ApJ, 333, 666 (IT88)

\bibitem{} Isobe T., Feigelson E.D., Akritas M.G., and Babu G.J., 1990, 
	Ap.J., 364, 104.

\bibitem{} Jannuzi, B.T., Smith, P.S., \& Elston, R., 1993, ApJ 428, 130 (J94) 

\bibitem{} Kinney A.L., Bohlin R.C., Blades J.C \& York D.G., 1991, ApJS 75, 645 (K91) 

\bibitem{} Lamer G., Brunner H. \& Staubert R., 1996, A\&A 311, 384 (L96) 

\bibitem{} Lanzetta K. M., Wolfe A.M. \& Turnshek D.A., 1995, ApJ 440, 435 (L95) 

\bibitem{} Maraschi L., Ghisellini G., Treves A. \& Tanzi E.G., 1986, APJ, 310, 325

\bibitem{} Matt G., Guainazzi M., Frontera F. et al., 1997, A\&A, 325, L13

\bibitem{} Mazzarella J.M. \& Balzano V.A., 1986, ApJS 62, 751 (MB86) 

\bibitem{} Mead A.R.G, Ballard K.R., Brand P.W.J.L, Hough J.H., Brindle C, 
	Bailey J.A., 1990, A\&AS 83, 183 (M90) 

\bibitem{} McNaron-Brown K. et al., 1995, ApJ, 451, 575 (MB95)

\bibitem{} Nass P., Bade N., Kollgaard R.I., Laurent--Muehleisen S.A.,
Reimers D. \& Voges W., 1996, A\&A 309, 419 (N96) 

\bibitem{} Orr A. et al., 1998, To appear in ``The Active X-ray Sky:
Results from BeppoSAX and Rossi-XTE'', Nuclear Physics B Proceedings
Supplements, L. Scarsi, H. Bradt, P. Giommi and F. Fiore (eds.), Elsevier
Science B.V

\bibitem{} Padovani P. \& Giommi P., 1995, APJ, 444, 567 

\bibitem{} Padovani P. \& Giommi P., 1995, MNRAS 277, 1477 (PG95) 

\bibitem{} Padovani P. \& Giommi P., 1996, MNRAS, 279, 526

\bibitem{} Padovani P., et al., 1998, To appear in ``The Active X-ray Sky:
Results from BeppoSAX and Rossi-XTE'', Nuclear Physics B Proceedings
Supplements, L. Scarsi, H. Bradt, P. Giommi and F. Fiore (eds.), Elsevier
Science B.V

\bibitem{} Parmar A.N., Martin D.D.E., Bavdaz M., et al., 
		1997, A\&AS 122, 309.

\bibitem{} Patnaik A. R., Browne I.W.A., Wilkinson P.N. \& Wrobel J.M.,
     1992, MNRAS 254, 655 (P92) 

\bibitem{} Perlman E.S., Stocke J.T., Schachter, J.F., Elvis, M.,
Ellingson, E., Urry, C.M., Potter M., Impey C.D., Kolchinsky, P. 1996a, 
APJS, 104, 251 (P96a) 

\bibitem{} Perlman E.S., E.S. Stocke J.T., Wang Q.D. Morris S.L., 
1996b, ApJ, 456, 451 (P96b) 

\bibitem{} Perlman E.S. \& Stocke J.T., 1993, ApJ 406, 430 (PS93) 

\bibitem{} Pesce J.E., Falomo R., Treves A., 1994, AJ 107, 494 (P94) 

\bibitem{} Pian E., et al., 1998, ApJ, 492, L17

\bibitem{} Pian E. \& Treves A., 1993, ApJ, 416, 130 (Pi93)

\bibitem{} Pian, E., Falomo, R., Scarpa, R., \& Treves, A., 1994, ApJ 432, 547 (Pi94) 

\bibitem{} Punch, M., et al., 1992, Nature, 358, 477

\bibitem{} Quinn J., et al., ApJ, 1996, 456, L83

\bibitem{} Raiteri C.M., Villata M., De Francesco G., Lanteri L., 
Cavallone M., Sobrito G., A\&AS, 1998, (in press).

\bibitem{} Remillard R.A., Tuohy I.R., Brissenden R.J.V., Buckely D.A.H., Schwartz D.A., 
   Feigelson E.D. \& Tapia S., 1989, ApJ 345, 140 (R89) 

\bibitem{} Sambruna R.M., Barr P., Giommi P., Maraschi L.,
      Tagliaferri G.,  Treves A., 1994, ApJS, 95, 371 (Sa94)

\bibitem{} Sambruna R.M., Maraschi L., and Urry C.M., 1996, ApJ, 463, 444.

\bibitem{} Smith P.S., Jannuzi B.T., Elston R., 1991, ApJS 77, 67

\bibitem{} Stecker O.C. \& De Jager F.W., 1997, ApJ, 476, 712


\bibitem{} Stocke J.T., Morris S.L., Gioia I.M., Maccacaro T.,
	Schild R.E. \& Wolter A., 1990, ApJ 348, 141 (S90) 

\bibitem{} Tovmassian H.M., Sherwood W.A., Sherwood V.E., Schultz G.V., Salter C.J. \&
     Matthews H.E., 1984, A\&AS 58, 317 (T84) 

\bibitem{} Ulrich, M.-H., Maraschi, L., and Urry, C.M., 1997, ARAA, n. 35.

\bibitem{} Villata M., Raiteri C.M., Ghisellini G., et al., 1997, 
	A\&AS 121, 119

\bibitem{} Villata M., Raiteri C.M., Lanteri L., Sobrito G., Cavallone M., 
	1998, A\&AS, in press.

\bibitem{} Wolter A., Caccianiga A., Della Ceca R., Maccacaro T., 1994,
     ApJ, 433, 29 (W94) 

\bibitem{} Worrall D.M., Wilkes B.J., 1990, ApJ, 360, 396 (WW90)

\end{thebibliography}
\end{document}